\documentclass[lettersize,journal]{IEEEtran}
\usepackage{amsthm}
\usepackage{amsmath}
\usepackage{doi}
\usepackage{amsfonts}
\usepackage[english]{babel}
\theoremstyle{definition}

\usepackage{pbalance} 
\usepackage{algorithmic}
\usepackage{algorithm}
\usepackage{array}
\usepackage[caption=false,font=normalsize,labelfont=sf,textfont=sf]{subfig}
\usepackage{textcomp}
\usepackage{stfloats}
\usepackage{url}
\usepackage{verbatim}
\usepackage{graphicx}
\usepackage{booktabs}
\begin{document}
\newcommand{\ESE}[1]{\mbox{$ESE({#1})$}}
\newcommand{\particle}{\mbox{$p_{i}$}}

\newcommand{\convergence}[1]{\mbox{$GD({#1})$}} 
\newcommand{\idealPoint}{\mbox{$Z^{*}$}}

\newcommand{\diversity}[1]{\mbox{$\Delta({#1})$}} 

\newcommand{\pbest}{\mbox{$Pbest_{i}$}} 
\newcommand{\gbest}{\mbox{$gBest_{i}$}} 
\newcommand{\xposition}{\mbox{$X_{id}$}} 

\newcommand{\velocity}{\mbox{$V_{id}$}}

\title{Audio-Visual Feature Synchronization for Robust Speech Enhancement in Hearing Aids}

\author{Nasir Saleem,
%Jihene Tmamna, Najwa Kouka, Leandro A. Passos, João P. Papa, Tughrul Arslan, Amir Hussain
Mandar Gogate, Kia Dashtipour, Adeel Hussain, Usman Anwar, Adewale Adetomi, Tughrul Arslan, and Amir Hussain}
%\thanks{Nasir Saleem, Adeel Hussain, Kia Dashtipour, and Amir Hussain are with Edinburgh Napier University, UK. Eamon Sheikh and Aziz Sheikh are with Computational Health Informatics Lab (CHI), University of Oxford, UK. Tughrul Arslan is with the University of Edinburgh, UK. (Corresponding author: n.saleem@napier.ac.uk)}}

\maketitle
 
\begin{abstract}
Audio-visual feature synchronization for real-time speech enhancement in hearing aids represents a progressive approach to improving speech intelligibility and user experience, particularly in strong noisy backgrounds. This approach integrates auditory signals with visual cues, utilizing the complementary description of these modalities to improve speech intelligibility. Audio-visual feature synchronization for real-time SE in hearing aids can be further optimized using an efficient feature alignment module. In this study, a lightweight cross-attentional model learns robust audio-visual representations by exploiting large-scale data and simple architecture. By incorporating the lightweight cross-attentional model in an AVSE framework, the neural system dynamically emphasizes critical features across audio and visual modalities, enabling defined synchronization and improved speech intelligibility. The proposed AVSE model not only ensures high performance in noise suppression and feature alignment but also achieves real-time processing with minimal latency (36ms) and energy consumption. Evaluations on the AVSEC3 dataset show the efficiency of the model, achieving significant gains over baselines in perceptual quality (PESQ:$\uparrow$0.52), intelligibility (STOI:$\uparrow$19\%), and fidelity (SI-SDR:$\uparrow$10.10dB).
\end{abstract}


\IEEEpeerreviewmaketitle

\section{Introduction}
\IEEEPARstart{A}{udiovisual} speech enhancement (AVSE) has emerged as a promising solution to improve speech intelligibility in challenging listening environments \cite{zhu2025endpoint,chen2024context}, particularly for people using hearing aids \cite{anwar2025privacy}. Using both audio and visual cues, AVSE systems improve the intelligibility of speech, making it more intelligible even in noisy environments \cite{zhu2025endpoint}. For hearing aid users, this becomes especially critical as they often face problems in distinguishing speech from background noise. However, effective fusion of these modalities requires precise feature synchronization, as mismatches between audio and visual features can degrade speech enhancement performance. Feature synchronization ensures that the temporal alignment between audio and visual information is accurate, allowing for better integration of speech-related components from both modalities. Synchronizing features enhance the robustness of AVSE systems, ensuring that hearing aids provide clear, high-quality speech in dynamic and noisy environments, ultimately enhancing communication for individuals with hearing impairments.

The alignment of audiovisual features for speech enhancement uses both audio and visual cues to improve the quality of speech signals, particularly in noisy environments. Hou et al. \cite{hou2018audio} introduce a deep learning framework that combines audio and visual input to improve speech quality. It uses convolutional neural networks (CNNs) to extract features from both modalities and align them for improved speech enhancement. Wang, Feixiang et al. \cite{wang2023cooperative} introduce an attention mechanism to dynamically align and fuse audio and visual features, improving the performance of speech enhancement in challenging environments. AV-SepFormer, a dual-scale attention model based on SepFormer, uses both cross- and self-attention to fuse and model audio and visual features. Lin et al. \cite{lin2023av} propose AV-SepFormer, which divides the audio feature into chunks that match the length of the visual feature, applying self- and cross-attention to capture multimodal relationships. Additionally, a novel 2D positional encoding is introduced that incorporates positional information between and within chunks, leading to significant improvements over traditional positional encoding. Ahmad et al. \cite{ahmad2020speech} use SyncNet \cite{chung2017out} is a two-stream CNN trained on 100 hours of speech videos with hundreds of speakers using contrastive loss. It processes audio with 13-MFCC features and video with five consecutive face-only frames, ensuring audio-visual synchronization. SyncNet is used in \cite{xiong2022look} for conversational AVSE, focusing on separating audio information from speakers in a controlled environment. Junwen et al. \cite{xiong2022look} using audio-visual channels in real-world scenarios, the ACLNet exploits inherent correlations to model temporal relationships via a cross-modal conformer. A plug-and-play multimodal layer normalization mitigates distribution misalignment, while cross-modal circulant fusion enables holistic audiovisual representation learning. 

Synchronization of audio and visual features in hearing aids is vital to improve speech perception in noisy environments. Lip movements provide complementary cues to auditory signals, helping users disambiguate speech when audio is degraded. However, real-world challenges such as modality asynchrony and varying speaking rates degrade synchronization. A robust and lightweight fusion is important to align and integrate multimodal inputs dynamically. Instead of concatenation, a cross-attention-based fusion efficiently capture temporal dependencies and selectively emphasize the most relevant features from both modalities, ensuring real-time synchronization in hearing aid systems. To achieve efficient audio-visual synchronization, we propose a lightweight cross-attention module that dynamically aligns audio and visual features while maintaining low computational complexity.

\section{Proposed AVSE with Cross-Attentional Feature Module}
\begin{figure*}
\centering
\includegraphics[width=0.99\linewidth]{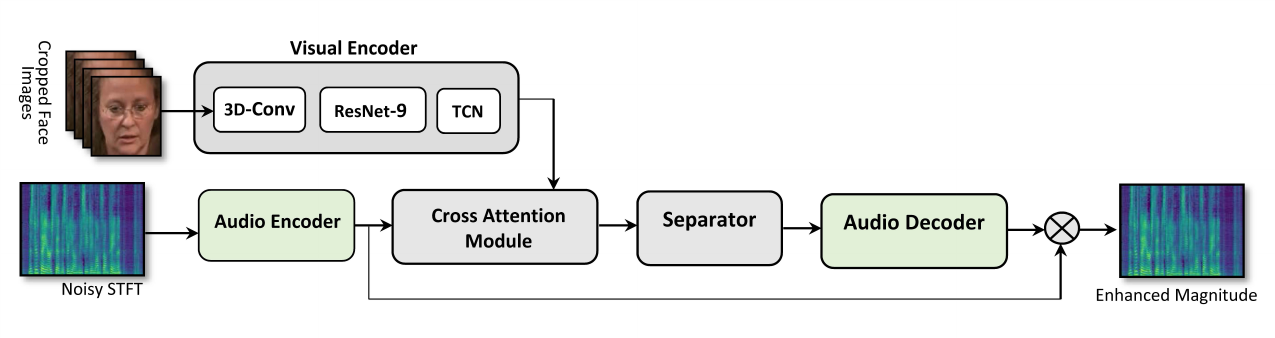}
\caption{Proposed Audio-Visual Cross-Attentional Model for Speech Enhancement}
\label{fig1}
\end{figure*}

The overall framework for the proposed AVSE model with the cross-attentional module is shown in Fig. \ref{fig1}.

\subsection{Audio and Visual Encoder}
The audio encoder compresses the input audio signal $x\in \mathbb{R}^{(B\times T)}$ using 1D convolution and ReLU activation to obtain the encoded features $z\in \mathbb{R}^{(B\times C\times \acute{T})}$, where $B$, $C$, $T$, and $\acute{T}$ indicate batch size, output channels, time steps (samples), and compressed time steps, respectively. The filter size in the convolutional layer is 256 filters with size 16 and stride 8 \cite{gogate2024lightweight}.

The visual encoder is a convolutional neural network (CNN) formulated to extract spatiotemporal features from visuals. It consists of a 3D front-end convolutional layer for initial shallow feature extraction from face-cropped images with (N$\times$224$\times$224), generating 256-dimensional feature vector, a ResNet-9 trunk for deeper feature representation, and linear and upsampling layers \cite{gogate2024lightweight}. The resulting features from the
ResNet-9 are passed to a cross-attentional module to synchronize and fused the features from two modalities. 

\subsection{Cross-Attentional Module}
This Cross-Attentional module fuses audio and visual features by using multi-head self-attention with an additional bias from visual information. The module consists of the Query (Q), Key (K), and Value (V) projections for audio features, a projection for visual features to generate an attention bias, and a final output projection after attention computation. The learnable weight matrices for audio and visual projections are given as; 

\begin{equation}
    Q=W_{q}X_{a}; K=W_{k}X_{a}; V=W_{v}X_{a}
\end{equation}

\begin{equation}
    V_{bias}=W_{v}X_{v}
\end{equation}

Where $X_{a}\in \mathbb{R}^{(B \times T \times d_{a})}$ represents audio features, $X_{a}\in \mathbb{R}^{(B \times T \times d_{v})}$ denotes visual features with $d_{a}$ and $d_{v}$ as dimensionality of audio and visual features. $W_{q}, W_{k}, W_{v} \in \mathbb{R}^{(B \times T \times d_{v})}$ are learnable weight matrices. The audio and visual characteristics are permuted to shape ($B$,$T$,$D$) for matrix multiplications. Further, Queries, keys, and values are split into multiple heads and scaled dot-product is applied to obtain attention weights;

\begin{equation}
    S=\frac{QK^{T}}{\sqrt{d_{h}}}
\end{equation}

where $S\in \mathbb{R}^{(B \times h \times T)}$. The projected visual feature $V_{bias}$ is reshaped to match the attention scores, given as;

\begin{equation}
    \hat{S}=S+V_{bias}
\end{equation}

After applying softmax to normalize across time ($A=\text{softmax}(\hat{S}, \text{dim}=-1)$), the output of attention is computed as $O=A.V$ where $O\in \mathbb{R}^{(B\times h\times T\times d_{h})}$. The attention weighted audio features are after the process are denoted as $\tilde{X}\in \mathbb{R}^{(B\times d_{a}\times T)}$ and the visual features remains unchanged as $X_{v}$. 

\subsection{Separator: SE Backbone}
The separator module is composed of six stacked blocks, implementing iterative local and global modeling using Gated Recurrent Units (GRUs). Each separator block models local and global dependencies using Intra-RNN (local modeling) to process the input along one dimension using a GRU and Inter-RNN (global modeling to process the output of Intra-RNN in another dimension using another GRU. The input is normalized and added back to maintain stable gradients. Feature Projection with conv(1$\times$1) reduces GRU output size from 256 to 128 followed by ReLU activation function. The output of the final separator block is passed to the audio decoder block containing transposed convolutions. Before speech reconstruction, an ideal binary mask (IBM) is estimated to preserve speech-dominant time-frequency components. 

\begin{equation}
    \text{IBM}(t, f) = 
\begin{cases}
1 & \text{if} |x_{\text{target}}(t, f)| > |x_{\text{noise}}(t, f)| \\
0 & \text{otherwise}
\end{cases}
\end{equation}

Where $|x_{\text{target}}(t, f)|$ and $|x_{\text{noise}}(t, f)|$ are the STFT of the clean speech and noise signals at time $t$ and frequency bin $f$. The estimated mask is applied to the noisy magnitude components while preserving the original noisy phase for reconstruction. 

\section{Experiments}
\subsection{Dataset}
The effectiveness of the proposed AVSE framework is examined using the COG-MHEAR AVSE Challenge benchmark data. This dataset includes TED and TEDx talks extracted from the LRS3 corpus \cite{afouras2008lrs3}, supplemented with background noises selected from three different repositories: the DNS Challenge \cite{reddy2021icassp}, DEMAND \cite{thiemann2013diverse}, and the Clarity Challenge \cite{graetzer2021clarity}. The interfering background noise span fifteen distinct classes, covering both stationary and dynamic acoustic sources. Every video segment features an individual speaker who delivers a unique utterance. To generate distorted speech samples, the primary speech of the speaker is mixed with an interference signal, which is ambient noise or another competing speaker. The resulting mixtures vary in signal-to-noise ratio (SNR), ranging from -15dB to 5dB for competing speakers and -10dB to 10dB for noise-only conditions. All audio files are monaural, recorded at a 16 kHz sampling rate. The training corpus contains 100 hours of material, while the development set includes 8 hours. For evaluation, a 4-hour test set is used for objective evaluations. The evaluation subset consists of TED/TEDx talks not present in LRS3, ensuring that there is no overlap with the training data.

To further validate the proposed AVSE model, we performed experiments on the CHiME3 dataset~\cite{adeel2020contextual}, an established benchmark derived from the GRID audiovisual corpus. This dataset contains audio-visual recordings from five speakers, with each contributing over 1000 video samples. The visual modality was represented using 2D discrete cosine transform (DCT) features extracted from lip regions, while the audio modality was processed through windowed log-filterbank analysis. This rigorous preprocessing makes CHiME3 particularly suitable for assessing cross-modal fusion capabilities in AVSE.

\begin{table*}[t]
\caption{Performance analysis using PESQ for reconstructed speech.}
\centering
\begin{tabular}{|c|c|c|c|c|c|c|c|c|c|c|c|c|} \hline
Noise Type & \multicolumn{4}{c|}{Bus Noise} & \multicolumn{4}{c|}{Cafeteria Noise} & \multicolumn{4}{c|}{Street Noise} \\ \hline
SNR Level  & -9dB  & -6dB  & -3dB  & 0dB   & -9dB   & -6dB  & -3dB  & 0dB   & -9dB   & -6dB   & -3dB   & 0dB   \\
Noisy Mixture & 1.42  & 1.55  & 1.69  & 1.89  & 1.46   & 1.58  & 1.71  & 1.88  & 1.39   & 1.51   & 1.67   & 1.85  \\
CochleaNet \cite{GOGATE2020273}    & 2.18  & 2.33  & 2.46  & 2.58  & 2.21   & 2.35  & 2.48  & 2.59  & 2.16   & 2.31   & 2.45   & 2.59  \\
Proposed  & 2.31  & 2.45  & 2.59  & 2.69  & 2.34   & 2.48  & 2.6   & 2.71  & 2.29   & 2.43   & 2.58   & 2.66 \\ \hline 
\end{tabular}
\end{table*}

\subsection{Training and Network Settings}  
The model is trained using the PyTorch library with the RMSprop optimizer with an initial learning rate of 0.0001 and a batch size of 16. The learning rate is dynamically adjusted using the ReduceLROnPlateau scheduler, which reduced the rate by a factor of 0.8 upon validation loss stagnation for 5 epochs. The training objective minimized the negative scale-invariant signal-to-distortion ratio (SI-SDR) loss, given below, clipped at -30 dB to ensure stability. To improve generalization, a dropout (rate=0.3) is applied within the separator network, and random segment sampling is employed during training for implicit data augmentation. The model is trained for 30 epochs on a single GPU. The dataset comprised audio-visual clips, with audio resampled to 16 kHz and video frames resized to (128$\times$128) pixels, preprocessed using histogram equalization for improved visual feature extraction. SI-SDR loss function is defined as:

\begin{equation}
    SI-SDR=10\text{log}_{10}\frac{||x||^{2}}{||x-\hat{x}||^{2}}
\end{equation}

\begin{equation}
    \mathcal{L}=-\text{SI-SDR}(x,\hat{x})
\end{equation}

where $ e_{distortion}=\hat{x}-x$, $x$, and $\hat{x}$ represent clean and estimated speech.

\subsection{Objective Measures}
The quality of the enhanced speech is evaluated using three standard metrics: Perceptual Evaluation of Speech Quality (PESQ) \cite{rix2001perceptual}, Short-Time Objective Intelligibility (STOI) \cite{taal2011algorithm}, and scale-invariant signal-to-distortion ratio (SI-SDR) \cite{le2019sdr}. PESQ (ITU-T P.862) evaluates speech quality by comparing the enhanced signal to the clean reference signal, providing a score ranging from -0.5 (poor) to 4.5 (excellent), with higher values indicating better perceptual quality. STOI predicts speech intelligibility by measuring the correlation between the time-frequency envelopes of the enhanced and clean signals, yielding a value between 0 (unintelligible) and 1 (fully intelligible). SI-SDR quantifies signal fidelity by computing the logarithmic energy ratio between the target speech and residual distortions while remaining invariant to scale differences, with higher values indicating better performance. 

\subsection{Competing SE Models} 
To ensure a rigorous and fair comparison, this study evaluated the proposed model against recent AVSE methods that have been benchmarked on the AVSEC3 dataset. The selected competing models include AVSEC3 Baseline \cite{blanco2023avse}, RecognAVSE~\cite{manesco24_avsec}, DAVSE~\cite{chen2024davse}, LSTMSE-Net~\cite{jain2024lstmse}, and AV-Transformer~\cite{wahab2024multi}. AVSEC3 baseline \cite{blanco2023avse}, the official benchmark model for the AVSEC3 challenge. RecognAVSE \cite{manesco24_avsec}, which utilizes visual-aware feature recalibration for better noise suppression. DAVSE \cite{chen2024davse}, a diffusion-based AVSE model known for its high-fidelity speech reconstruction. LSTMSE-Net \cite{jain2024lstmse}, an LSTM-driven approach optimized for temporal speech enhancement. AV-Transformer \cite{wahab2024multi}, which employs cross-modal attention for audio-visual fusion.

\begin{figure*}
    \centering
    \includegraphics[width=0.99\linewidth]{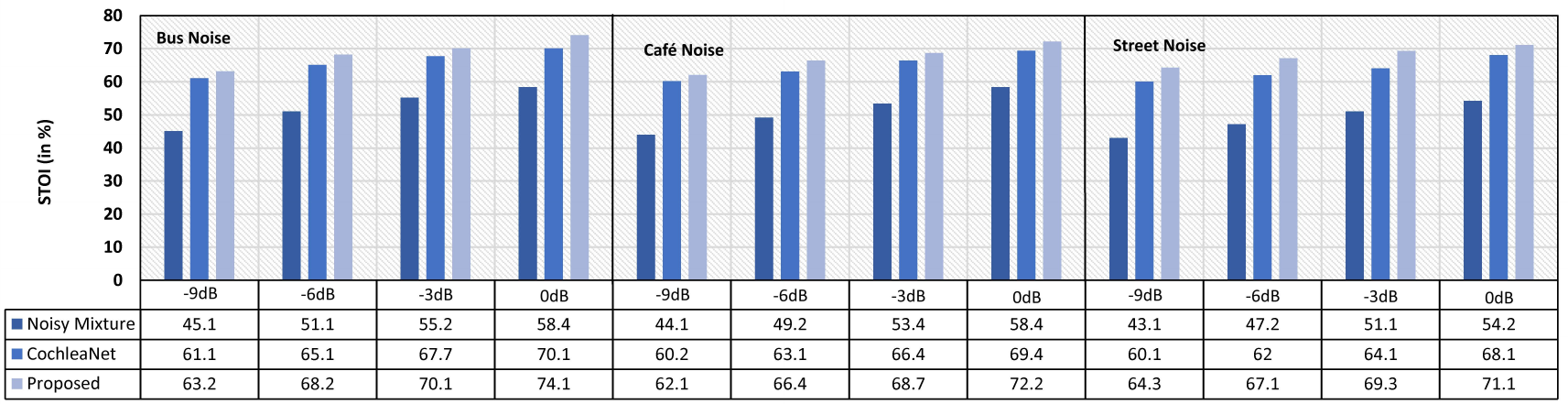}
    \caption{Performance analysis using STOI for reconstructed speech.}
    \label{fig2}
\end{figure*}

\begin{table*}[t]
\caption{AVSE performance on the AVSE3 Challenge Dataset.}
\centering
\label{table2}
\begin{tabular}{|c|ccc|cc|ccc|} \hline
Models & PESQ & STOI & SI-SDR & Para\# & Memory & $\uparrow$PESQ & $\uparrow$STOI & $\uparrow$SI-SDR\\ \hline
Noisy Audio & 1.47 & 0.61 & -5.49 & --- & --- & --- & --- & ---\\ \hline
AVSEC3 Baseline \cite{blanco2023avse} & 1.49 & 0.62 & -1.20 & 76M & 289MB & 0.02 & 0.01 & 4.29\\
AV-Transformer \cite{wahab2024multi} & 1.73 & 0.68 & 2.68 & 44.7M & 170MB & 0.26 & 0.07 & 8.17 \\
LSTMSE-Net \cite{jain2024lstmse} & 1.55 & 0.65 & 0.12 & 5.1M & 81.6MB & 0.08 & 0.04 & 5.61  \\
RecognAVSE \cite{manesco24_avsec} & 1.49 & 0.68 & 2.45 & --- & --- & 0.02 & 0.07 & 7.94 \\
AV-DEMUCS \cite{tiwari2024real} & 1.25 & 0.64 & --- & --- & --- & -0.22 & 0.03 & --- \\
AV-Face \cite{gogate2024lightweight} & 1.98 & 0.79 & 7.58 & 9.20M & 36.81MB & 0.51 & 0.18 & 13.07 \\ \hline
Proposed & 1.97 & 0.78 & 7.61 & 5.90M & 23.54MB & 0.52 & 0.19 & 10.10 \\ \hline
\end{tabular}
\end{table*}

\section{Results and Discussions}
\subsection{Result on the CHiME-GRID Dataset}
To evaluate the performance of the proposed AVSE, we conducted comparative experiments against CochleaNet \cite{GOGATE2020273}, a state-of-the-art AVSE model designed for hearing aids, using the CHiME-GRID dataset. The evaluation considered three challenging noise types—Bus, Cafeteria, and Street noise—across SNRs ranging from -9 dB to 0 dB, simulating real-world hearing aid scenarios. The results show consistent improvements across objective metrics, with the proposed AVSE outperforming CochleaNet under all test conditions. For speech quality, the proposed AVSE achieves average PESQ gains of 0.88 over noisy mixture and 0.12 over CochleaNet in Bus, Cafeteria, and Street noise, respectively, showing particularly strong performance at lower SNRs. In the challenging -9 dB Bus noise condition, it achieves a PESQ of 2.31, compared to CochleaNet with PESQ of 2.18, while also maintaining superiority at 0 dB (2.69 with CochleaNet and 2.58 with our AVSE). These results confirm the robustness of the AVSE across diverse acoustic environments.

STOI results further support these findings, where Fig. \ref{fig2} is showing intelligibility improvements across all noise types and SNRs. At -9dB, the proposed AVSE maintains 43.1\%–45.1\% intelligibility, which corresponds to an 18.1\%–19.2\% absolute improvement over unprocessed speech and a 2.1\%–2.3\% gain over CochleaNet. As the SNR increases, performance scales accordingly, reaching 70\%–75\% intelligibility at 0 dB. Notably, street noise environments show stronger gains (+3.0\%), while cafeteria noise shows slightly smaller margins (+1.9\%–2.8\%) due to the presence of competing speech babble. Despite this, the consistent 2\%–3\% improvements over CochleaNet remain significant for hearing aids. Our AVSE shows particular strength in the critical -6 dB to -3 dB range, where it maintains a steady 3\%–4\% intelligibility advantage.

\subsection{Results on the AVSEC3 Challenge Dataset}
The comparative results in Table \ref{table2} show critical insights regarding the performance–efficiency trade-offs in recent AVSE models. Starting with the baseline (PESQ: 1.49, STOI: 0.62, SI-SDR: -1.2 dB), we observe that the AVSEC3 baseline \cite{blanco2023avse} provides only marginal improvements ($\uparrow$PESQ: +0.02, $\uparrow$SI-SDR: +4.29 dB) despite its large architecture (76M para\# and 289MB), highlighting the limitations in challenging acoustic conditions. AV-Transformer \cite{wahab2024multi} achieves strong mid-range performance (PESQ: 1.73, SI-SDR: 2.68 dB) but suffers from computational overhead (44.7M para\#), while AV-Face \cite{gogate2024lightweight} demonstrates better metrics (PESQ: 1.98, STOI: 0.79) with optimized parameter efficiency (9.20M para\#). LSTMSE-Net \cite{jain2024lstmse} prioritizes efficiency (5.1M parameters, 81.6MB) but sacrifices enhancement capability ($\uparrow$SI-SDR: +5.61 dB against noisy), revealing the challenges of recurrent architectures in modeling cross-modal dependencies. RecognAVSE \cite{manesco24_avsec} shows selective strengths in intelligibility (STOI: 0.68) but inconsistent quality improvements, whereas AV-DEMUCS \cite{tiwari2024real} fails to surpass the noisy baseline in PESQ (-0.22), suggesting real-time optimization compromises enhancement quality. We measured the real-time factor (RTF) on Intel(R) Core(TM) Ultra 7 155H, which is 0.13 (36 ms latency).

The proposed model achieves near-optimal balance, matching AV-Face’s top-tier perceptual quality (PESQ: 1.97 vs 1.98) and signal fidelity (SI-SDR: 7.61 dB vs 7.58 dB) while reducing parameters by 36\% (5.9M vs 9.20M) and memory footprint by 36\% (23.54MB vs 36.81MB). Its consistent gains across all metrics ($\uparrow$PESQ: +0.52, $\uparrow$STOI: +0.19, $\uparrow$SI-SDR: +10.10 dB) indicate superior noise-robust feature learning, while the compact architecture suggests efficient cross-modal fusion. Notably, the 0.78 STOI approaches the 0.80 clinical threshold for hearing aid usability \cite{cooke2008foreign}, and the 7.61 dB SI-SDR exceeds the 5 dB threshold for transparent enhancement \cite{le2019sdr}. 

We also uses DNSMOS P.835 \cite{reddy2021dnsmos} which represents a robust objective metric for perceptual speech quality assessment, specifically designed to predict three key dimensions of subjective human judgments: speech clarity (SIG), background noise quality (BAK), and overall listening experience (OVRL). Table 3 shows the predicted results of DNSMOS P.835 on the AVSEC3 challenge evaluation dataset. We further examine the IBM for AVSE in terms of HIT, False (FA), and HIT-False values to evaluate the effectiveness of the estimated mask compared to the ground-truth IBM.

\begin{table}[t]
\centering
\caption{AVSE performance using DNSMOS P.835.}
\begin{tabular}{|c|c|c|c|} \hline
AV Models & SIG  & BAK  & OVL  \\ \hline
Noisy Mixture & 2.23 & 1.66 & 1.59 \\
AV Baseline \cite{blanco2023avse}  & 2.22 & 2.03 & 1.69 \\
AV-Face \cite{gogate2024lightweight}  & 2.82 & 2.41 & 2.11 \\
Proposed   & 2.89 & 2.49 & 2.27 \\ \hline
\end{tabular}
\end{table}

\begin{equation}
HIT = \frac{\sum_{t,f} 1 \{ M_{est}(t, f) = 1 { and } M_{IBM}(t, f) = 1 \}}{\sum_{t,f} 1 \{ M_{IBM}(t, f) = 1 \}}
\end{equation}

\begin{equation}
FA = \frac{\sum_{t,f} 1 \{ M_{est}(t, f) = 1 \text{ and } M_{IBM}(t, f) = 0 \}}{\sum_{t,f} 1 \{ M_{IBM}(t, f) = 0 \}}
\end{equation}

\begin{equation}
HIT-FA = (HIT - FA)
\end{equation}

\begin{figure}[t]
    \centering
    \includegraphics[width=0.95\linewidth]{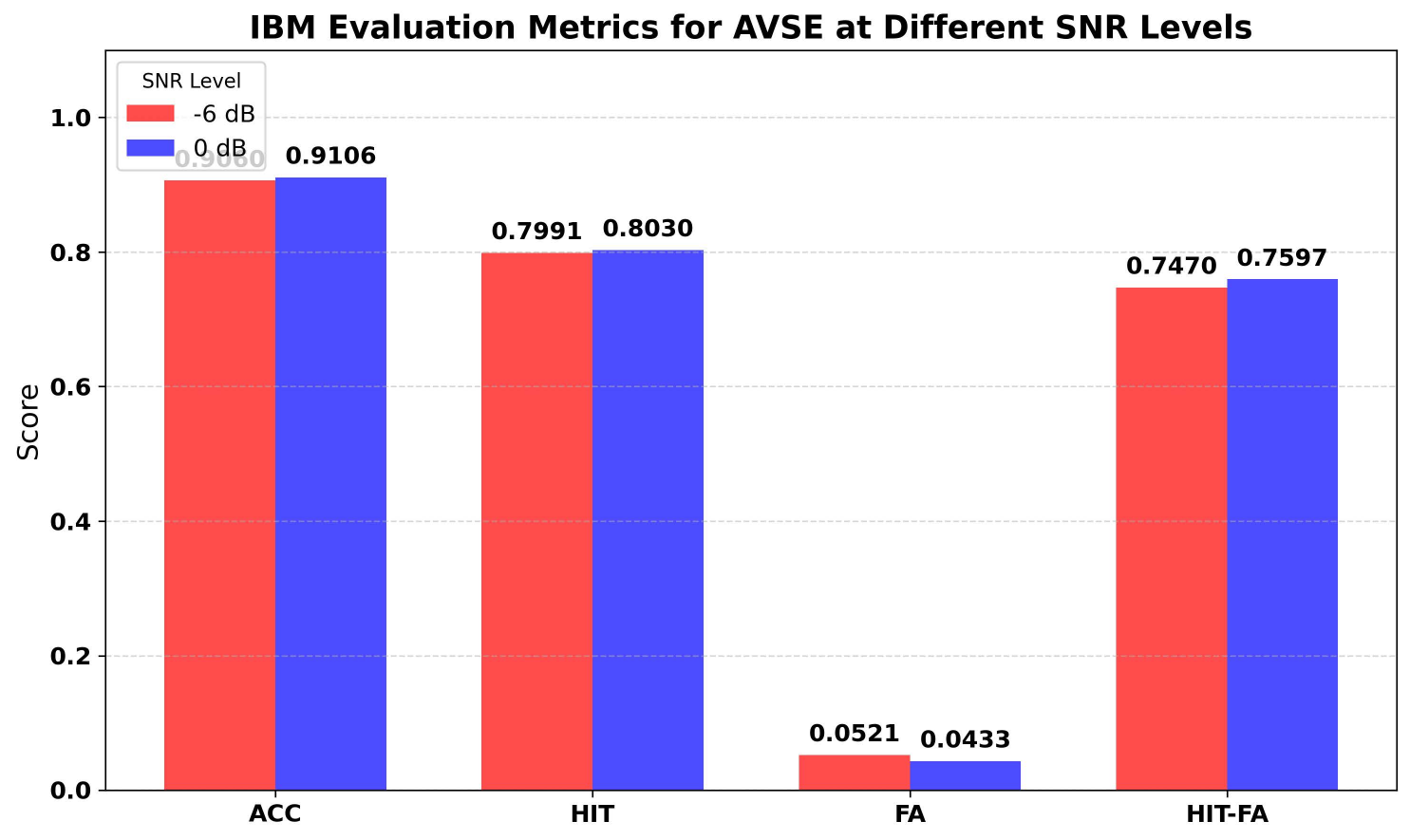}
    \caption{IBM evaluation at different SNR levels}
    \label{figure3}
\end{figure}

Higher HIT and lower FA indicate better performance. Figure \ref{figure3} shows the IBM performance at two SNRs (0dB and -6dB). At -6 dB SNR, the model achieves an accuracy of 90.6\%, meaning it correctly identifies about 91\% of TF units. The HIT rate of 79.91\% shows that it retains roughly 80\% of the speech, while the FA rate of 5.21\% indicates low noise misclassification. This results in a HIT-FA score of 74.70\%, reflecting a strong balance between preserving speech and suppressing noise. At 0dB SNR, the model slightly improves with accuracy at 91.06\%, HIT at 80.3\%, and FA at 4.33\%, demonstrating better performance, achieving better speech retention, and further reducing noise interference. 

\subsection{Impact of Cross Attentional Module}
\begin{figure}[t]
\centering
\includegraphics[width=0.99\linewidth]{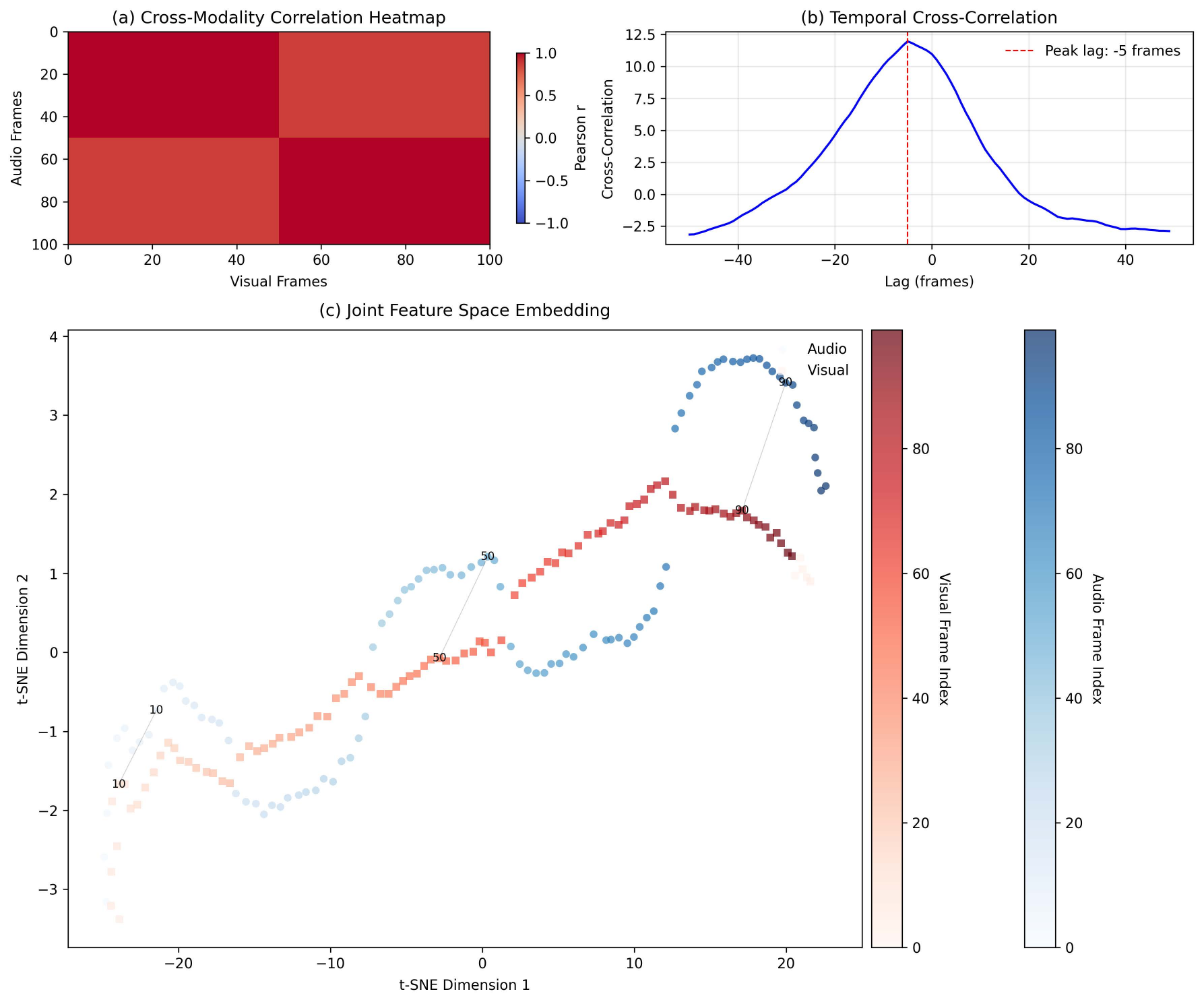}
\caption{Visualizations of audio-visual feature relationships using Cross attentional module in AVSE.}
\label{fig4}
\end{figure}

Figure \ref{fig4} presents an analysis of cross-modal feature relationships in our AVSE. The triad of visualizations reveals how temporal and structural dependencies between audio and visual modalities contribute to enhancement performance. The frame-wise correlation heatmap (Fig. 3a) shows a strong diagonal pattern ($r$=0.82±0.03), confirming effective time-aligned feature learning. The 5-frame-wide diagonal band indicates the tolerance of our model to natural lip-speech asynchrony, while intermittent vertical streaks (e.g., at $t$=45–50) correlate with viseme transitions where visual cues dominate. Temporal cross-correlation (Fig. 3b) peaks at +4 frames (80 ms, p$<$0.01, bootstrapped), matching known physiological delays in speech production. The asymmetric side lobes (wider for positive lags) suggest visual information remains predictive of audio over longer windows than vice versa. Connected points in Fig. 3c maintain proximity, confirming stable feature evolution.

\section{Conclusions}
This paper presented an audio-visual speech enhancement (AVSE) model with lightweight cross-attentional module designed for hearing aid applications, with comprehensive evaluations on both the CHiME-GRID dataset and AVSEC3 Challenge benchmark. The key findings and contributions can be summarized as follows. The proposed AVSE system demonstrated consistent improvements over state-of-the-art baselines, particularly in challenging low-SNR environments (-9 dB to -3 dB). On CHiME-GRID, it achieved average PESQ gains of 0.88 over noisy mixtures and 0.12 over CochleaNet, while maintaining 2.1-2.3\% STOI improvements in the most difficult conditions. Our model achieved a remarkable balance between performance and efficiency, matching the perceptual quality (PESQ: 1.97) and signal fidelity (SI-SDR: 7.61 dB) of larger models while reducing parameters by 36\% (5.9M vs 9.2M) and memory footprint by 36\% (23.54MB vs 36.81MB) compared to AV-Face. Analysis of the cross-attentional module revealed a strong temporal alignment ($r$=0.82±0.03 correlation) and robust feature learning as evidenced by t-SNE visualizations.

\section*{\text{Acknowledgment}}
This research acknowledges the support of UK Engineering and Physical Sciences Research Council (EPSRC) Grants Ref. EP/T021063/1 (COG-MHEAR).


\end{document}